# Intershell correlations in endohedral atoms

**M. Ya. Amusia**[a, b]* and **L.V. Chernysheva**[b]

[a]Racah Institute of Physics, the Hebrew University, Jerusalem, Israel;
[b]A. F. Ioffe Physical-Technical Institute, St. Petersburg, Russian Federation

We have calculated partial contributions of different endohedral and atomic subshells to the total dipole sum rule in the frame of the random phase approximation with exchange (RPAE) and found that they are essentially different from the numbers of electrons in respective subshells. This difference manifests the strength of the intershell interaction. We present concrete results of calculations for endohedrals, composed of fullerene $C_{60}$ and all noble gases He, Ne, Ar, Kr and Xe thus forming respectively He@$C_{60}$, Ne@$C_{60}$, Ar@$C_{60}$, Kr@$C_{60}$, and Xe@$C_{60}$. For comparison we obtained similar results for isolated noble gas atoms. The deviation from number of electrons in outer subshells proved to be much bigger in endohedrals than in isolated atoms thus demonstrating considerably stronger intershell correlations there.

**Introductory remarks**

In this paper we investigate the role of intershell correlations in atoms and endohedrals, i.e. fullerenes staffed by some other atom. Investigation of electron correlations in isolated atoms since long ago was and still is an important area of studies in the physics of atoms. As correlation we consider the effects of that part of the inter-electron interaction which is neglected in the frame of a crude picture of an atom that treats atomic electrons as moving independently, each one in the field of the nucleus and the so-called self-consistent field created by averaged action of all other electrons. The most accurate choice of the one-electron field is the Hartree-Fock approximation (HF) (see, e.g. [1] and references therein).

To take into account the electron correlations we will use the so-called random phase approximation with exchange (RPAE) that gives particularly good results in photoionization calculations that is the objects of this research [1, 2]. In RPAE frame it is natural to distinguish intra- and intershell correlations. The former include interaction of electrons that belong to a given atomic subshell while the latter are results on interaction of electrons that belong to different subshells (or shells). These correlations in the physics of atoms are well reflected in the absolute and differential cross section of photoionization of atomic subshells and result in formation of prominent resonances in the partial photoionization cross-sections [2], e.g. the interference resonances [3]. Recently it was disclosed that intershell correlations in atoms are strongly reflected in the so-called partial dipole sum-rules [4].

The aim of this paper is to demonstrate that the influence of a very big (as compared to atomic) fullerene electron shell affect very strong the partial photoionization cross-sections of an encapsulated atom and this effect is prominently reflected in the partial sum-rules of this atom. These modifications of the partial sum rules demonstrate directly the importance of the intershell interaction in photoionization of endohedrals that is dominated by the interaction between atomic and fullerenes electrons. Thus, we intend to demonstrate that by studying the dipole sum rule that correspond to a given subshell of an endohedral one obtains important information on the interaction between electrons of this subshell and the fullerenes electrons.

W. Thomas, F. Reiche, and W. Kuhn have discovered the Dipole sum rule almost a century ago [5, 6]. It connects oscillator strengths of discrete $f_k$ transitions, dipole non-relativistic photoabsorption cross-section $\sigma(\omega)$ as a function of incoming photon frequency $\omega$ and the number of electrons $N$ in the system under consideration, e.g. atom or endohedral that absorb photons, by the following relation (see e.g. [2])[1]:

$$\sum_{\text{All } k} f_k + \frac{c}{2\pi^2} \int_I^\infty \sigma(\omega) d\omega \equiv S = N. \quad (1)$$

Here $I$ is the ionization potential and $c$ is the speed of light. The relation (1) is a strict theoretical statement in non-relativistic approximation, but experimentally observed cross-sections and oscillator strengths has admixtures of non-dipole contributions and of a whole variety of relativistic corrections. These corrections grow rapidly with increase of $\omega$ and become dominative in $\sigma(\omega)$ at relativistic energy $\omega \geq c^2$. Fortunately, for most of the atoms as well as endohedrals photon energies $I \leq \omega \ll c^2$ saturate the integral in (1), so the role of dipole contributions is most important. To determine experimentally the absolute values of the photoionization cross-section is a hard task, so instead very often one normalizes the accurately measured relative values of $\sigma(\omega)$ using (1). This is a regular procedure not only for atoms, but also for more complex objects, e.g. such, as endohedrals and fullerenes [7].

It exist a widely distributed believe that an equation, similar to (1), although approximate, is valid for partial subshell contributions, at least for isolated atoms [1]:

$$\sum_{\text{All } k_i} f_{k_i} + \frac{c}{2\pi^2} \int_{I_i}^\infty \sigma_i(\omega) d\omega \equiv S_i \approx N_i. \quad (2)$$

Here the oscillator strength $f_{k_i}, \sigma_i(\omega), I_i$ and $N_i$ are, respectively, the discrete excitations oscillator strengths, photoionization cross-section, ionization potential and total number of electrons in the $i^{\text{th}}$ subshell. A usual assumption is that relation (2) is accurate enough even to attribute absolute values to the measured relative partial cross-sections.

Known almost half a century, the random phase approximation with exchange (RPAE) gives very good results in description of partial photoionization cross-sections, angular distributions of photoelectrons and spin polarization parameters of isolated atoms [1]. An interesting feature of this approximation is the fact that the dipole sum rule (1) is precisely valid in its frame. Recently, however, we investigate the partial sum rules in RPAE frame and have demonstrated that, unexpectedly, for multi-electron $f$ and $d$ subshells of isolated atoms the values $S_f$ and $S_d$ (see (2)) are considerably bigger than the respective $N_f = 14$ and $N_d = 10$. It means that for other subshells the inequalities hold $S_{s,p} < N_{s,p}$ where $N_s = 2$ and $N_p = 6$ [2], in generally holds [4].

Here we perform calculations for noble gas endohedrals, in which noble gas atoms are stuffed inside an almost ideally spherical fullerene $C_{60}$. We employ the one-electron Hartree-Fock (HF) approximation, in both length and velocity forms of the operator that describes the photon-electron interaction, denoted by upper indexes $L$ and $\nabla$. We use HF equations properly modified to take into account the additional potential that reproduces the fullerenes shell (see, e.g. [8] and references therein). In HF the relation (1) is essentially violated. We take into account the multi-electron correlations using RPAE generalized in such a way that permits to take into account the static potential of the fullerenes shell and its dynamic polarization that affects the intensity of the photon beam that causes the inner atom photoionization.

---

[1] We employ the atomic system of units $m = e = \hbar = 1$. Here $m$ is the electron mass; $e$ is its charge and $\hbar$ is the Planck constant.
[2] All considered here atoms have only closed subshells.

Note that in RPAE frame the Eq. (1) is valid. In this sense, it exist a fundamental difference between ordinary and endohedral atoms. For ordinary atoms Eq. (1) has to be fulfilled, while for endohedrals – not, since endohedral is a subsystem of the total system that consist of atom A and fullerene $C_N$, with its total number of electron equal to $N = N_A + N_{C_N}$, where $N_A$ and $N_{C_N}$ are the total number of atomic and endohedral electrons, respectively.

**Essential formulas**
The necessary details about HF and RPAE equation and their solutions one can find in [9, 10]. Here we present only important definitions and the main points of calculation procedures. The HF equation for an endohedral looks like

$$-\frac{\Delta}{2}\phi_j(x) - \frac{Z}{r}\phi_j(x) + U_C(r) + \sum_{k=1}^{N_A}\int \phi_k^*(x')\frac{dx'}{|\mathbf{r}-\mathbf{r}'|}\left[\phi_k(x')\phi_j(x) - \phi_j(x')\phi_k(x)\right] = E_j\phi_j(x) \quad (3)$$

Here Z is the nuclear charge, $U_C(r)$ is the fullerenes static potential in the frame of its spherical model, $\phi_j(x)$ is the one-electron wave function, $x \equiv \vec{r},\vec{\sigma}$ are the electron coordinate and spin variables, $E_j$ is the one-electron HF energy; the summation is performed over all occupied electron states $N_A$ of the atom A, stuffed inside the fullerene $C_N$. Eq. (3) differs from that for an isolated atom by addition of endohedral potential $U_C(r)$.

The oscillator strength is determined by the square module of dipole matrix elements in the length $\vec{\varepsilon}\vec{r}$ or velocity $\vec{\varepsilon}\vec{\nabla}$ forms, calculated between HF wave functions (3) of the electron that undergoes transition from the initial state $i$ to the final $f$ due to photon absorption:

$$d_{if}^L = \omega_{if}\int \phi_i^*(x)(\vec{\varepsilon}\vec{r})\phi_f(x)dx, \quad d_{if}^\nabla = \int \phi_i^*(x)(\vec{\varepsilon}\vec{\nabla})\phi_f(x)dx, \quad \omega_{if} \equiv E_f - E_i \quad (4)$$

The following expression determines the oscillator strength of a one-electron transition $i \to f$:

$$f_{if}^{L,\nabla} = \frac{2}{\omega_{if}}\left|d_{if}^{L,\nabla}\right|^2, \quad (5)$$

Similar to (5) expression is valid for continuous spectrum excitations that is connected to the photoionization cross-section of the $i$ subshell by the following relation

$$\sigma_i^{L,\nabla}(\omega) = \frac{2\pi^2}{c} f_{iE_{(i)}}^{L,\nabla}, \quad E_{(i)} = \omega - I_i. \quad (6)$$

Relations similar to (5) and (6) give the oscillator strengths and photoionization cross-section in RPAE, if one substitutes the HF matrix elements $d_{if}^{L,\nabla}$ by solutions of RPAE equations

$$\langle i|D(\omega)|f\rangle = \langle i|d^{L,\nabla}|f\rangle + \left(\sum_{v\leq F, v'>F} - \sum_{v>F, v'\leq F}\right)\frac{\langle v'|D(\omega)|v\rangle\langle vi|V|v'f - fv'\rangle}{[\omega - E_{v'} + E_v \pm i\delta)]}. \quad (7)$$

Here $V$ denotes the Coulomb inter-electron interaction, sums over $v \leq F$ include occupied one-electron states, while sums over $v > F$ include excited discrete levels and integration over continuous excitation energies. In the denominator the sign $\pm$ means + for $v'$ vacant and – for $v'$ occupied one-electron states, respectively. Note that $D$ does not have $L, \nabla$ indexes, since in RPAE corresponding values are equal [2, 1].

Along with static fullerenes potential $U_C(r)$ in (3), in the frame of RPAE for endohedrals one has to include the dipole polarization of the fullerene electron shell by the incoming photon beam. This hard problem can be considerably simplified assuming validity of the following strong inequalities: $R_C \gg r_A$ and $R_C \gg \Delta$, where $R_C, r_A, \Delta$ are the fullerene and atomic radiuses and the thickness of the fullerene shell, respectively. In reality, one has instead $R_C > r_A$ and $R_C > \Delta$. However, for reasonable accounting of the polarization of the fullerenes shell influence, we assume that strong inequalities hold, so that $r_A / R_C \ll 1$ and $\Delta / R_C \ll 1$.

Retaining only lowest order terms in expansion of the endohedral photoionization amplitude $D_{if}^{A@C_N}(\omega)$ in powers of $r_A / R_C \ll 1$ and $\Delta / R_C \ll 1$, one obtains

$$D_{if}^{A@C_N}(\omega) \cong \left[1 - \frac{\alpha_C(\omega)}{R_C^2}\right]\langle i|D(\omega)|f\rangle \equiv G(\omega)\langle i|D(\omega)|f\rangle \equiv G(\omega)D_{if}(\omega). \quad (8)$$

Here $\alpha_C(\omega)$ is the dynamic dipole polarizability of the fullerene. Eq. (6) determines the A@C$_{60}$ photoionization cross-section after substituting (8) into (5) instead of $d_{if}^{L,\nabla}$.

An important step of calculations is the choice of the endohedral static potential $U_C(r)$ shape [11, 8]. Our preference is the Lorentz shape [11, 8] because it is free from serious minuses of the very often used in studies of endohedral photoionization square well potential [12]:

$$U_C(r) = -U_o R_C \frac{d}{(r - R_C)^2 + d^2}. \quad (9)$$

Here $d$ is the potential width.

The choice of parameters in (9) is the following: $d = d_C / 2 \cong 1.46$, $R = 6.72$; $U_0$ leads to the electron affinity of $C_{60}^-$ that coincide with its experimental value [11]. The function $\alpha_C(\omega)$ for (8) one can find in [1], page 307.

**Results of calculations**
We performed calculations using computing codes system ATOM-M [10]. Tables 1-5 collect the results for total $S$ and partial $S_i$ sums, defined by (1) and (2). The results for He@C$_{60}$ and He (Table 1) demonstrate an important role on fullerene shell even in He case. Note that for He $S^{RPAE} > 2$ and $S_{He}^{RPAE} < 2$. Table 2 collect results for Ne@C$_{60}$ and Ne. Only outer subshell increases its contribution by more than a factor of 2. Table 3 presents results for Ar@C$_{60}$ and Ar. The contribution of 1s, 2s, and 3s almost completely disappears. The 2p-subshell is a loser (-0.65) in Ar@$C_{60}$, but a winner in Ar (+0.99). The biggest winner is 3p. Table 4 demonstrates the results for Kr@$C_{60}$ and Kr. Again, contributions of s-subshells decrease impressively, particularly that of 4s. As in Ar case, 2p-subshell is a loser (-1) in Kr@$C_{60}$, but a winner in Kr (+1.01). 3p is a loser (~-2.6), while 3d is a big winner (+4.7). The biggest winner, with (+29!) is 4p in Kr. Table 5 presents the results for Xe@$C_{60}$ and Xe. As in other cases, all s-subshells are losers. The same is for p-subshells, except the outer 5p that for Xe@$C_{60}$ is a very big winner (+25.7). As to 3d and 4d, for Xe@$C_{60}$ and Xe they are winners, acquiring from +2.31 for 3d in Xe to 5.3 for 4d in Xe@$C_{60}$.

Our calculations show that all partial sums $S_i$ are essentially different from the number of electrons in the respective subshell $N_i$. The difference takes place not only in RPAE but even stronger in HF, thus signalling redistributing of oscillator strength already on the one-electron HF level. Comparison with isolated atoms demonstrates that the presence of the fullerene shell considerably enhances, particularly for outer subshells, the deviation from the number of electrons in the respective subshells.

Note that after performing summation over all $i$, we obtain $S^{RPAE} > N_A$ for endohedrals and $S_A^{RPAE} < N_A$ for atoms A, respectively. The difference $S^{RPAE} - N_A$ reaches 25.3 in Kr@$C_{60}$ and then start to decrease becoming 20.6 in Xe@$C_{60}$. The inequality $S^{RPAE} > N_A$ demonstrates that the atom A increases its $S$-value due to interaction with the fullerenes shell. This is achieved in spite of the fact that according to the data in the Tables all inner atomic subshells are losers, while only the outer subshells with $l \geq 1$ are winners. The big increase of the contribution from outer subshells is a result of existence in these subshells so-called Giant endohedral resonances [13]. Fullerene shell increases the loses $\Delta_i$ in inner subshells of Ne and Ar leaving them almost unaltered in Kr and Xe since their inner subshells correspond to bigger binding energies than in Ne and Ar.

The inequality $S_A^{RPAE} < N_A$ instead of equality $S_A^{RPAE} = N_A$ that follows from RPAE theory reflects the defects of our calculation procedures. The differences $N - S_A^{RPAE} > 0$ characterize the contributions to (1) of the cross-sections long "tails" that are beyond upper limits of our numeric integrations and of neglected discrete excitation levels, since we include only four of them for each subshell. However, it is hard to imagine that these calculation peculiarities could essentially affect the redistribution of the partial $S_i$ in endohedrals that is the main result of this paper.

In most of the considered cases, the more electrons has a subshell, the bigger is the surplus $\Delta_i = S_i - N_i$ that goes from low-electron subshells due to intershell interaction. The very fact that this redistribution is a manifestation of the intershell interaction is easy to understand. Indeed, in absence of this interaction relation $S_{i,A}^{RPAE} = N_i$ becomes valid for any subshell. For equal numbers of $N_i$ of a given endohedral or atom, $\Delta_i$ increases with growth of the principal quantum number.

In studies of multi-electron atoms that require complex numeric calculations it is hard to explain results pure qualitatively, since the interplay of a number of different tendencies affect them. It is known, however, that the cross-sections of subshells with small angular momentum $l = 0;1$, decreases with $\omega$ growth much slower than that with $l = 2$. The account of intershell interaction increases at high $\omega$ the cross sections of $l = 2$ subshells prominently [14]. In addition, one has to note that the interelectron interaction role is bigger when the relative role of nuclear charge is smaller, i.e. in outer subshells with many electrons and big principal quantum numbers. The big number of electrons in the fullerene shell and concentration of its photoionization cross-section at relatively small $\omega$ explains the very big increase of C$_{60}$ upon $S_{outer}$. For them $\Delta_i$ are the biggest, as Tables 1-5 demonstrate.

**Discussion and conclusions**
By studying the partial sum rules, we demonstrated here that intershell interaction affects prominently the absolute photoionization cross-section of endohedrals in a very broad frequency range that is reflected on the partial sum rule level. It demonstrates previously unobserved overall sufficiently strong intershell or interference interaction in both atoms and endohedrals. Before, such interaction was considered as a rather specific feature that is effective in relatively narrow $\omega$ regions only in few outer shells, and near so-called Giant resonances of the intermediate subshells [3, 15]. No doubt that similar to presented here is the situation for many other endohedrals and isolated atoms.

It would be very interesting to perform experimental investigation aiming to demonstrate the prominent violation of the partial sum rules. This is not a simple task, having in mind that for each subshell $i$ the measurements must be performed in a broad $\omega$ region in coincidence with creation of only $i$ vacancy. However, such an experiment would be of great importance for the understanding of electronic structure of endohedrals, as well as isolated atoms.

Table 1. Partial and total sums $S_{1s}^{RPAE}, \Delta_{1s} \equiv S_{1s}^{RPAE} - 2$ for He@C$_{60}$ and He ($N=Z=2$). Superscript $A$ denotes atomic values.

| He | S/s | $N$ | $S_{1s}^{RPAE}$ | $\Delta_{1s}$ | $S_{1s}^{RPAE}$ | $S_{1s,A}^{RPAE}$ | $\Delta_{1s}$ |
|---|---|---|---|---|---|---|---|
| 1 | 1s | 2 | 1.06 |  | 3.93 | 1.64 | 0.8 |

Table 2. Partial and total sums $S_{i,HF}^{L}, S_{i,HF}^{\nabla}, S_{i}^{RPAE}, \Delta_i \equiv S_i^{RPAE} - N_i$, and $\sum_{\leq i} S_i^{RPAE}$ for Ne@C$_{60}$ and Ne ($N=Z=10$). Superscript $A$ denotes atomic values. S/s means subshell

| Ne | S/s $i$ | $Ni$ | $S_{i,HF}^{L}$ | $S_{i,HF}^{\nabla}$ | $S_i^{RPAE}$ | $\Delta_i$ | $\sum_{\leq i} S_i^{RPAE}$ | $S_{i,A}^{RPAE}$ | $\Delta_{i,A}$ | $\sum_{\leq i} S_{i,A}^{RPAE}$ |
|---|---|---|---|---|---|---|---|---|---|---|
| 1 | 1s | 2 | 1.06 | 1.06 | 1.06 | -0.94 | 1.06 | 1.4 | -0.6 | 1.4 |
| 2 | 2s | 2 | 0.85 | 0.76 | 0.51 | -1.24 | 1.57 | 0.72 | -1.28 | 2.12 |
| 3 | 2p | 6 | 14.4 | 10.3 | 12.8 | +6.8 | 14.4 | 7.27 | +1.27 | 9.39 |

Table 3. Partial and total sums $S_{i,HF}^{L}, S_{i,HF}^{\nabla}, S_{i}^{RPAE}, \Delta_i \equiv S_i^{RPAE} - N_i$, and $\sum_{\leq i} S_i^{RPAE}$ for Ar@C$_{60}$ and Ar ($N=Z=18$). Superscript $A$ denotes atomic values. S/s means subshell

| Ar | S/s $i$ | $Ni$ | $S_{i,HF}^{L}$ | $S_{i,HF}^{\nabla}$ | $S_i^{RPAE}$ | $\Delta_i$ | $\sum_{\leq i} S_i^{RPAE}$ | $S_{i,A}^{RPAE}$ | $\Delta_{i,A}$ | $\sum_{\leq i} S_{i,A}^{RPAE}$ |
|---|---|---|---|---|---|---|---|---|---|---|
| 1 | 1s | 2 | 0.28 | 0.28 | 0.28 | -1.72 | 0.28 | 0.8 | -1.2 | 0.8 |
| 2 | 2s | 2 | 0.57 | 0.57 | 0.57 | -1.43 | 0.85 | 1.00 | -1.0 | 1.80 |
| 3 | 2p | 6 | 5.57 | 4.74 | 5.35 | -0.65 | 6.2 | 6.99 | +0.99 | 8.79 |
| 4 | 3s | 2 | 0.62 | 0.51 | 0.41 | -1.59 | 6.61 | 0.46 | -1.54 | 9.25 |
| 5 | 3p | 6 | 43.4 | 22.2 | 31.9 | +25.8 | 38.46 | 7.8 | +1.8 | 17.1 |

Table 4. Partial and total sums $S_{i,HF}^{L}, S_{i,HF}^{\nabla}, S_{i}^{RPAE}, \Delta_i \equiv S_i^{RPAE} - N_i$, and $\sum_{\leq i} S_i^{RPAE}$ for Kr@C$_{60}$ and Kr ($N=Z=36$). Superscript $A$ denotes atomic values. S/s means subshell

| Kr | S/s $i$ | $Ni$ | $S_{i,HF}^{L}$ | $S_{i,HF}^{\nabla}$ | $S_i^{RPAE}$ | $\Delta_i$ | $\sum_{\leq i} S_i^{RPAE}$ | $S_{i,A}^{RPAE}$ | $\Delta_{i,A}$ | $\sum_{\leq i} S_{i,A}^{RPAE}$ |
|---|---|---|---|---|---|---|---|---|---|---|
| 1 | 1s | 2 | 0.85 | 0.85 | 0.85 | -1.15 | 0.85 | 0.84 | -1.16 | 0.84 |
| 2 | 2s | 2 | 1.19 | 1.19 | 1.19 | -0.81 | 2.04 | 1.20 | -0.80 | 2.04 |
| 3 | 2p | 6 | 5.07 | 4.73 | 5.0 | -1 | 7.04 | 4.99 | +1.01 | 7.03 |
| 4 | 3s | 2 | 0.91 | 0.83 | 0.73 | -1.27 | 7.77 | 0.73 | -1.27 | 7.76 |
| 5 | 3p | 6 | 3.68 | 3.43 | 3.43 | -2.57 | 11.20 | 3.44 | -2.56 | 11.2 |
| 6 | 3d | 10 | 16.6 | 12.5 | 14.7 | +4.7 | 25.91 | 14.66 | 4.66 | 25.86 |
| 7 | 4s | 2 | 0.45 | 0.36 | 0.42 | -1.58 | 26.33 | 0.38 | -1.62 | 26.24 |
| 8 | 4p | 6 | 49.9 | 24.7 | 35.0 | +29 | 61.32 | 7.29 | +1.29 | 33.53 |

Table 5. Partial and total sums $S_{i,HF}^{L}$, $S_{i,HF}^{\nabla}$, $S_{i}^{RPAE}$, $\Delta_i \equiv S_i^{RPAE} - N_i$, and $\sum_{\leq i} S_i^{RPAE}$ for Xe@C$_{60}$ and Xe ($N=Z=54$). Superscript $A$ denotes atomic values. S/s means subshell

| Xe | S/s $i$ | $N_i$ | $S_{i,HF}^{L}$ | $S_{i,HF}^{\nabla}$ | $S_i^{RPAE}$ | $\Delta_i$ | $\sum_{\leq i} S_i^{RPAE}$ | $S_{i,A}^{RPAE}$ | $\Delta_{i,A}$ | $\sum_{\leq i} S_{i,A}^{RPAE}$ |
|---|---|---|---|---|---|---|---|---|---|---|
| 1 | 1s | 2 | 0.5 | 0.5 | 0.5 | -1.5 | 0.5 | 0.5 | -1.5 | 0.5 |
| 2 | 2s | 2 | 0.93 | 0.93 | 0.93 | -1.07 | 1.43 | 0.93 | -1.07 | 1.43 |
| 3 | 2p | 6 | 3.58 | 3.58 | 3.58 | -2.42 | 5.01 | 3.57 | -2.43 | 5.00 |
| 4 | 3s | 2 | 0.93 | 0.93 | 0.93 | -1.07 | 5.94 | 0.92 | -1.08 | 5.92 |
| 5 | 3p | 6 | 4.47 | 4.08 | 4.3 | -1.7 | 10.24 | 4.17 | -1.83 | 10.09 |
| 6 | 3d | 10 | 14.8 | 12.6 | 13.9 | +3.9 | 24.12 | 12.31 | +2.31 | 22.4 |
| 7 | 4s | 2 | 0.74 | 0.65 | 0.65 | -0.35 | 24.77 | 0.68 | -1.32 | 23.08 |
| 8 | 4p | 6 | 2.38 | 2.12 | 2.24 | -3.76 | 27.01 | 2.28 | -3.72 | 25.36 |
| 9 | 4d | 10 | 19.3 | 12.4 | 15.3 | +5.3 | 42.31 | 14.82 | +4.82 | 40.18 |
| 10 | 5s | 2 | 0.58 | 0.37 | 0.63 | -1.37 | 42.94 | 0.53 | -1.47 | 40.71 |
| 11 | 5p | 6 | 55.0 | 24.7 | 31.7 | +25.7 | 74.61 | 8.73 | +2.73 | 49.44 |